\documentclass[twocolumn]{revtex4-1}

\usepackage{amsmath}    
\usepackage{natbib}
\newcommand{\Eq}[1]{Eq.~(\ref{#1})}
\newcommand{\Sec}[1]{Sec.~\ref{#1}}

\newcommand{\Fig}[1]{Fig.~\ref{#1}}
\newcommand{\Tab}[1]{Table~\ref{#1}}
\newcommand{\Ref}[1]{Ref.~\cite{#1}}

\usepackage{graphicx}   
\usepackage{verbatim}   
\usepackage{color}      
\usepackage{hyperref}   
\newcommand{\be}{\begin{equation}}
\newcommand{\ee}{\end{equation}}
\newcommand{\bi}{\begin{itemize}}
\newcommand{\ei}{\end{itemize}}

\begin{document}

\title{On the Feasibility and Utility of ISR Tagging}
\author{David Krohn}
\email{dkrohn@physics.harvard.edu}
\affiliation{Department of Physics, Harvard University, Cambridge MA, 02138}
\author{Lisa Randall}
\email{randall@physics.harvard.edu}
\affiliation{Department of Physics, Harvard University, Cambridge MA, 02138}
\author{Lian-Tao Wang}
\email{lianwang@princeton.edu}
\affiliation{Department of Physics, Princeton University, Princeton NJ, 08544}
\date{\today}
\begin{abstract}
{The production of new particles at a hadron collider like the LHC is always accompanied by   QCD 
radiation attributable to the
 initial state (i.e. ISR).
  This tends to complicate analyses, so ISR is normally regarded as a nuisance.  Nevertheless, we 
  show that ISR can also be valuable, yielding information that can help in the discovery and interpretation of physics beyond the Standard Model.
  To access this information we will  introduce new techniques designed to
identify ISR jets on an event-by-event basis, a process we term ISR tagging.  As a demonstration of their utility, we  will apply 
these techniques to  SUSY  di-squark (di-gluino) production to show that they can be used to identify ISR jets in roughly $40\%$ ($15\%$) of the events, 
with a mistag rate of around $10\%$ ($15\%$).  We then show that, through the application of a new method which we will introduce, 
knowledge of an ISR jet  allows us to infer the squark (gluino) mass to within roughly $20\%$ of its true value.
}
\end{abstract}
\maketitle

\section{Introduction}
Quarks and gluons are always splitting apart and recombining~\cite{Altarelli:1977zs}. When they are scattered at high energies this process is interrupted, and the result is that additional quarks and gluons which could not be recombined end up  in the final state.  This radiation, attributable to the splitting of the incoming states,
is termed initial state radiation (ISR).

ISR often complicates analyses. For example, it can overlap with, and thus contaminate, 
jets  formed from the decay of new particles (we will call these {\it FSR jets}). Furthermore, 
  when   ISR emissions yield additional independent jets (i.e. {\it ISR jets}), sorting out the combinatorics of an event can be even more difficult.  These are not irresolvable difficulties, and indeed recently progress has been made toward removing sources of  jet contamination~\cite{Butterworth:2008iy,*Ellis:2009su,*Ellis:2009me,*Krohn:2009th}, mitigating combinatorial difficulties~\cite{Alwall:2009zu}, defining new observables less sensitive to contamination~\cite{Nojiri:2010mk,*Konar:2010ma}, and more accurately accounting for the physics of ISR in Monte Carlo simulations~\cite{Alwall:2007fs,*Alwall:2010cq,*Plehn:2005cq,*Alwall:2008qv}. 

 However, ISR -- rather than always proving to be an obstacle -- can actually be helpful in the study of physics beyond the SM.
This has already been demonstrated in several recent studies.  In \Ref{Alwall:2008ve,*Alwall:2008va}, ISR was shown to make BSM events more prominent by giving pair-produced new-physics states something to recoil against, thus increasing both $\displaystyle{\not} E_T$ 
and $H_T$.  Meanwhile, \Ref{Papaefstathiou:2009hp,*Papaefstathiou:2010ru} took a different approach, calculating the influence of ISR on inclusive variables (which sum over everything measured in the detector) with the aim of inferring the mass scale probed in new-physics events.  

ISR's utility lies primarily in the fact that 
the physics of its production is (approximately) dependent only upon the mass scale of the event and the states it couples to -- independently of any intermediate or final decay products.  So while the kinematic quantities measured of  FSR jets in BSM  processes depend upon some combination of all the masses involved in the process, the kinematics of ISR depend upon {\it only} the masses of the particles involved in the production process (i.e. not upon those produced subsequently in a cascade/decay) and their couplings to initial state particles.

Here we will make use of this information carried by  ISR, with the main new feature of our approach being that rather than treating ISR in an inclusive way, as in \Ref{Papaefstathiou:2009hp,Papaefstathiou:2010ru}, {\it we will seek to identify a particular jet as attributable to ISR}.  We will see that such jets can be identified on an event-by-event basis with a small mis-tagging probability.  We will then show that this sort of technique for tagging ISR jets provides a powerful tool for understanding BSM physics that can be applied to many interesting processes. We focus in particular on a new kinematic method that can be used to evaluate squark and gluino masses.

The outline of this paper is as follows.  \Sec{sec:isrtag} will discuss the sort of techniques one can use to tag ISR jets,
\Sec{sec:isruse} will discuss one example of what we can    learn about BSM physics through the  study of ISR jets based solely on kinematics, and \Sec{sec:example} will apply these ideas to di-squark and di-gluinio production as a demonstration of their feasibility and utility.
Finally,  \Sec{sec:conclusions}  contains our conclusions.
\section{Tagging an ISR Jet}
\label{sec:isrtag}
Tagging an ISR jet requires identifying characteristics
that distinguish it from the other jets in an event.  Although ISR possesses 
some general traits that hold true regardless of the process at hand~\footnote{e.g. it is not very correlated in rapidity with FSR jets.}, tagging ISR jets solely based on these properties would be challenging, if not impossible. We instead  focus on tagging ISR in  a particular class of interesting processes -- the pair production of BSM particles, each of which decays into jets and an invisible particle (i.e. $pp\rightarrow N_fJ+2\chi_1^0+{\rm ISR}$ where $N_f=2/4$ for di-squark/di-gluino production).  Although we will restrict ourself to these topologies, we expect that ISR jets are also identifiable in other cases and that similar techniques could be developed for more complicated processes.  Nonetheless, this will serve as a proof of concept, that we employ later in \Sec{sec:example}, in what is already an important application of these ideas to BSM physics. 

It turns out that the modest assumption of pair production gives  one a significant handle for identifying ISR jets.  Suppose one expects to see $N_f$ FSR jets in a BSM event.  As long as there's no reason for these
to be particularly soft, one can assume that of the $N_f+1$ hardest jets in the event, $N_f$ are attributable to FSR and one to ISR.   
As the production process is symmetric, all of the properties governing the production of one FSR  jet should hold for the others.  Thus, of the $N_f+1$ hardest jets in the event, we can identify
the ISR jet as the one which is in some way distinguished from the others.

The method we prescribe for accomplishing this is to consider the $N_f+1$ hardest jets in an event and identify a candidate ISR jet (here labeled $i$) 
for which at least one of the following conditions is met~\footnote{It is possible, although rare, for more than one jet to pass Eqs.(\ref{eq:pttag}-\ref{eq:delta}).  If this happens we rank 
the tagging criteria by the order in which they are listed, e.g. accepting a jet which passes \Eq{eq:pttag} over one which passes only \Eq{eq:delta}, etc.  If more than one jet pass the same tagging 
criteria, we take as our candidate for  \Eq{eq:pttag} the hardest jet, and for \Eq{eq:delta} the jet with the largest $\Delta$.  If more than one jet pass \Eq{eq:raptag}, none are accepted.}:
\begin{enumerate}
\item The jet's $p_T$ is distinct (i.e. it is harder or softer than the others):
\be
\label{eq:pttag}
\frac{\max(p_{T i},p_{Tj})}{\min(p_{T i},p_{Tj})}> 2\ \forall\ j\neq i
\ee 
\item The jet is separated from the others in rapidity:
\be
\label{eq:raptag}
|y_i-y_j|>1.5\ \forall\ j\neq i
\ee
\item The jet is distinguished by its $m_i/p_{Ti}\equiv \Delta_i$ ratio~\footnote{We note that $\Delta$ 
is sensitive to higher order emissions, in contrast to the criteria of Eqs.(\ref{eq:pttag}-\ref{eq:raptag}).
Further work along this line, particularly in employing jet substructure (see \Ref{Abdesselam:2010pt,*Salam:2009jx} for a review), could potentially improve upon the methods presented herein.}:
\be
\label{eq:delta}
\frac{\max (\Delta_i,\Delta_j)}{\min(\Delta_i,\Delta_j)}> 1.5\ \forall\ j\neq i
\ee  
\end {enumerate} 
If a jet (again labeled $i$) is selected by any of the above criteria it should then satisfy {\it all} of the following:
\begin{itemize}
\item The selected jet must not be central: $|y_i| > 1$.
\item It must not be too close to the other jets, which are all implicitly FSR jets:
\be
|y_i-y_j|>0.5\ \forall\ j\neq i
\ee
\item These other jets must be 
reasonably close to each other in $p_T$:
\be
\label{eq:fsrptsim}
\frac{p_{Tj}}{p_{Tk}}<\rho+\frac{1/2}{1-\alpha}
\ee
for $p_{Tj(k)}=\max (\min)\{p_{Tl}|\forall\ l\neq i\}$, with $\rho=2(3)$ for $N_f=2(4)$,  and 
where we have introduced the variable 
\be
\alpha=\frac{\min(p_{T i},\displaystyle{\not} E_T)}{\max(p_{T i},\displaystyle{\not} E_T)}
\ee
to relax this condition when the ISR is very hard.  
\item Finally, the implicit FSR jets must be somewhat central: $|y_j|<2\ \forall\ j\neq i$
\end {itemize} 
If any of the above conditions is not satisfied, the jet being considered is not tagged and 
other jets are checked to see if they pass any of the distinguishing criteria (Eqs. \ref{eq:pttag}-\ref{eq:delta}).

We note that it is surely possible to improve upon the technique presented above, 
and that the numerical values we presented have  not been thoroughly optimized.  Even so, we will see these criteria already work quite well,
triggering on $40\%\ (15\%)$ of the events, for $N_f=2(4)$ topologies, with a small $10\%\ (15\%)$ mistag rate.
\section{Uses of an ISR Jet}
\label {sec:isruse}

Once an ISR jet has been identified in an event it can be used  in multiple ways to shed light upon the 
underlying physics that produced it.
As the production of ISR is determined by the mass scale probed by the process, the identity of the partons in the initial state, and 
the relevant parton distribution functions (PDFs), the resulting ISR kinematical distributions will reflect all of these influences~\footnote{
We will pursue these ideas in a followup paper.}.  
Here though, rather than focus on general properties of the aforementioned distributions, whose calculation would depend upon a careful treatment of QCD, we will instead present 
a simple new kinematical technique useful in measuring $m_{\rm BSM}$, the center of mass energy for the 
two heaviest BSM particles produced in the symmetric processes we are considering.  Because hadron colliders
tend to produce heavy states close to threshold, a measurement of  $m_{\rm BSM}$ is nearly equivalent to a measurement of the 
new-physics particle's mass: $m_{\rm BSM}=\sqrt{(p_{\tilde{q}/\tilde{g}}+p_{\tilde{q}^\ast/\tilde{g}})^2}\approx 2m_{\tilde{q}/\tilde{g}}$.

Other kinematic variables are also sensitive in some way to $m_{\rm BSM}$.  Examples include $M_{\rm eff}$~\cite{Hinchliffe:1996iu}, $M_{T_2}$~\cite{Lester:1999tx,*Barr:2003rg}, and their more advanced extensions~\cite{Lester:2007fq,*Cho:2007qv,*Cheng:2008mg}.  Some recent works~\cite{Konar:2009wn,*Cohen:2010wv} have also made use of ISR to give their $M_{T_2}$ 
distributions additional structure.  However, these techniques are in general sensitive to all of the masses in the decay chain, or only work for very specific processes (e.g. gluino stransverse mass~\cite{Cho:2007qv}).

Remarkably, by looking to ISR we can construct a new kinematic measure sensitive only to $m_{\rm BSM}$, independent of any other assumptions on the spectrum.
The basic idea behind this method stems from the observation that any BSM particles produced must be recoiling against ISR in the transverse plane.  
Boosting the FSR system back along the transverse plane
to compensate for the ISR jet's $p_T$ requires an assumption for the system's  center of mass energy, and only when we have assumed the correct
value will the boost function properly.  Before proceeding, we note that while any BSM particles are, in fact, recoiling against all of the ISR particles (rather than only the leading 
jet), in practice the ISR jets assume a strong $p_T$ hierarchy and using only the leading jet to apply a boost will serve as a reasonable approximation.

In detail, the method we prescribe to measure $m_{\rm BSM}$ using the ISR jet's kinematics is to:
\begin{enumerate}
\item Identify all of the visible FSR jets and boost them along the $z$ direction so that the visible FSR is at rest in the $z$ frame (i.e. the net $p_z$ for FSR jets is zero).  That is, each FSR four-vector (here labeled $i$) is shifted
\be
\label{eq:zboost}
E_i\rightarrow \gamma\left(E_i+\beta p_{iz}\right), \ p_{iz}\rightarrow \gamma (\beta E_i+p_{iz})
\ee
where $\beta=-p_z/E$ and $\gamma=1/\sqrt{1-\beta^2}$, for $p_z$ and $E$ the sum longitudinal momentum and energy taken over all observable particles in the system.
This boost is performed because, while ideally the system will be at rest in the $z$ direction before boosting in the transverse plane (step two), 
this is a configuration we cannot achieve because of uncertainties introduced by missing energy.
However, by applying the boost in \Eq{eq:zboost} we  approximate this condition.
\item Boost the system along the direction transverse to the beam, parallel to the transverse momentum of the ISR jet,  assuming some system mass $M$.  This means that the projection of each FSR $p_T$ vector along the ISR direction transforms as
\be
{p}_{\bar Ti}\rightarrow\frac{{p}_{T}^{\rm ISR}}{M}E_i+\sqrt{1+\left(\frac{{p}_{T}^{\rm ISR}}{M}\right)^2} {p}_{\bar Ti}
\ee
where ${p}_{\bar Ti}=\vec{p}_T\cdot \hat{p}_T^{\rm ISR}$ is the projection of each $p_T$ along the ISR $p_T$  direction.  
\item Measure the sum projection of the resulting boosted FSR along the ISR transverse direction, assigning 
the result a $\pm1$ depending upon the sign:
\be
\label {eq:sigma}
\sigma=
\begin{cases}
+1 &{\rm if \ }\sum_i{p}_{\bar Ti}>0 \\ 
-1 & {\rm if \ }\sum_i{p}_{\bar Ti}<0
\end{cases}
\ee
\item Finally, the average projection across many events is measured:
$\langle\sigma\rangle=\sum_{i=1}^N \sigma_i/N$
\end{enumerate}
When $\langle\sigma\rangle$ is positive there is a net projection along the ISR axis, indicating the assumed mass is too small, 
while when it is negative, the assumed mass is too large.  Examples of the resulting distributions are shown in \Fig{fig:imb}
for the case of di-squark and di-gluino production.

Before proceeding, we call attention to two choices we made in the analysis that might be improved in a more careful treatment.  
The first is in step one, where we boosted the FSR along the $z$ direction to approximate the longitudinal rest frame.  
While this technique seems to operate reasonably well, it may be possible to better infer the $z$-boost using beam thrust techniques
as  suggested in \Ref{Stewart:2010tn}.  We further note that \Eq{eq:sigma} assigns each event an equal weight when computing $\langle\sigma\rangle$, 
regardless of the measured imbalance.  This choice was made because weighting events by their $p_T$ imbalance ($\sum_i{p}_{\bar Ti}$)
 tends to make $\langle\sigma\rangle$ sensitive to only a few outlier events.  Perhaps a better measure exists, but we do not pursue it here.

\section{Example: Di-squark \& Di-gluino Production}
\label{sec:example}
We now apply the aforementioned techniques to the pair production of squarks and gluinos, letting $\tilde{q}\rightarrow q+\chi_1^0$ and $\tilde{g}\rightarrow q\bar{q}+\chi_1^0$ (via an off-shell squark).  To perform this analysis we  use \texttt{Madgraph v4.4.51}~\cite{Alwall:2007st} to generate $10^5$-event samples at matrix-element level, assuming a $14~\rm TeV$ LHC, which are then showered in \texttt{Pythia v6.422}~\cite{Sjostrand:2006za} and matched using the MLM procedure~\cite{Hoche:2006ph}.  Fully showered and hadronized events are then grouped into $0.1\times 0.1$ cells $(\eta, \phi)$ cells between $-5<\eta<5$, which are clustered in \texttt{Fastjet v2.4.2}~\cite{Cacciari:Fastjet,*Cacciari:2005hq} using the anti-$k_T$ algorithm~\cite{Cacciari:2008gp}.  Our di-squark samples were clustered using $R=0.7$, while $R=0.4$ was used for the busier di-gluino events.  Note that, to simplify matters, we have not accounted for the effects of multiple interactions or pileup. 

\Tab{tab:isreffics} shows the efficiencies found using the tagging procedure of \Sec{sec:isrtag}.  Remarkably, we see that the tagging efficiency (i.e. the percent of events in which an ISR jet is tagged) 
and the mistag rate (the percent of events in which a jet that has been tagged as ISR was tagged incorrectly) are stable, even when comparing a standard SUSY spectrum with $m_{\rm LSP}=100~\rm GeV$ 
to one in which the LSP is nearly degenerate with the supersymmetric particle that decayed into it.

\begin{figure*}
\begin{center}
\includegraphics[scale=0.22]{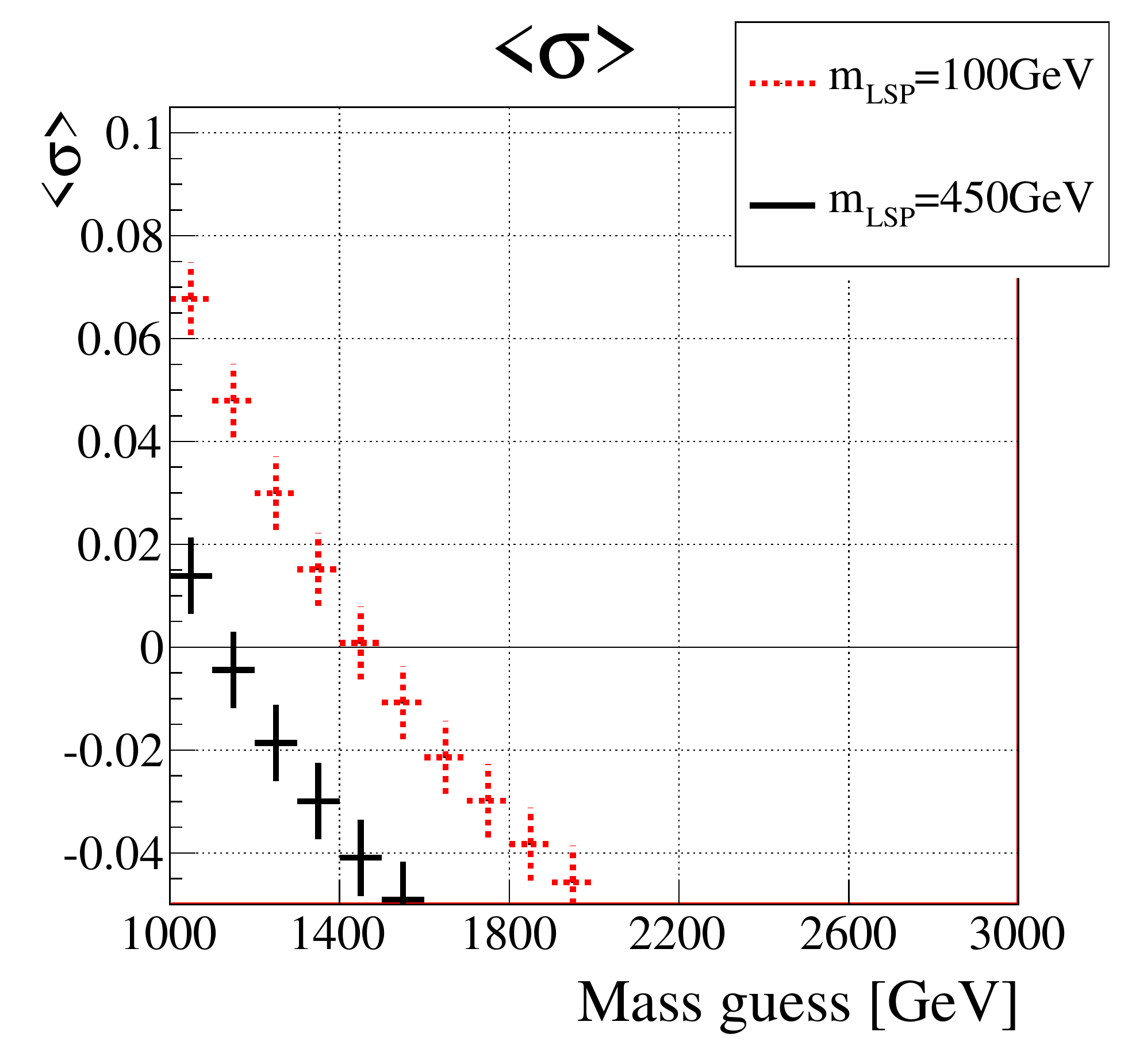}
\includegraphics[scale=0.22]{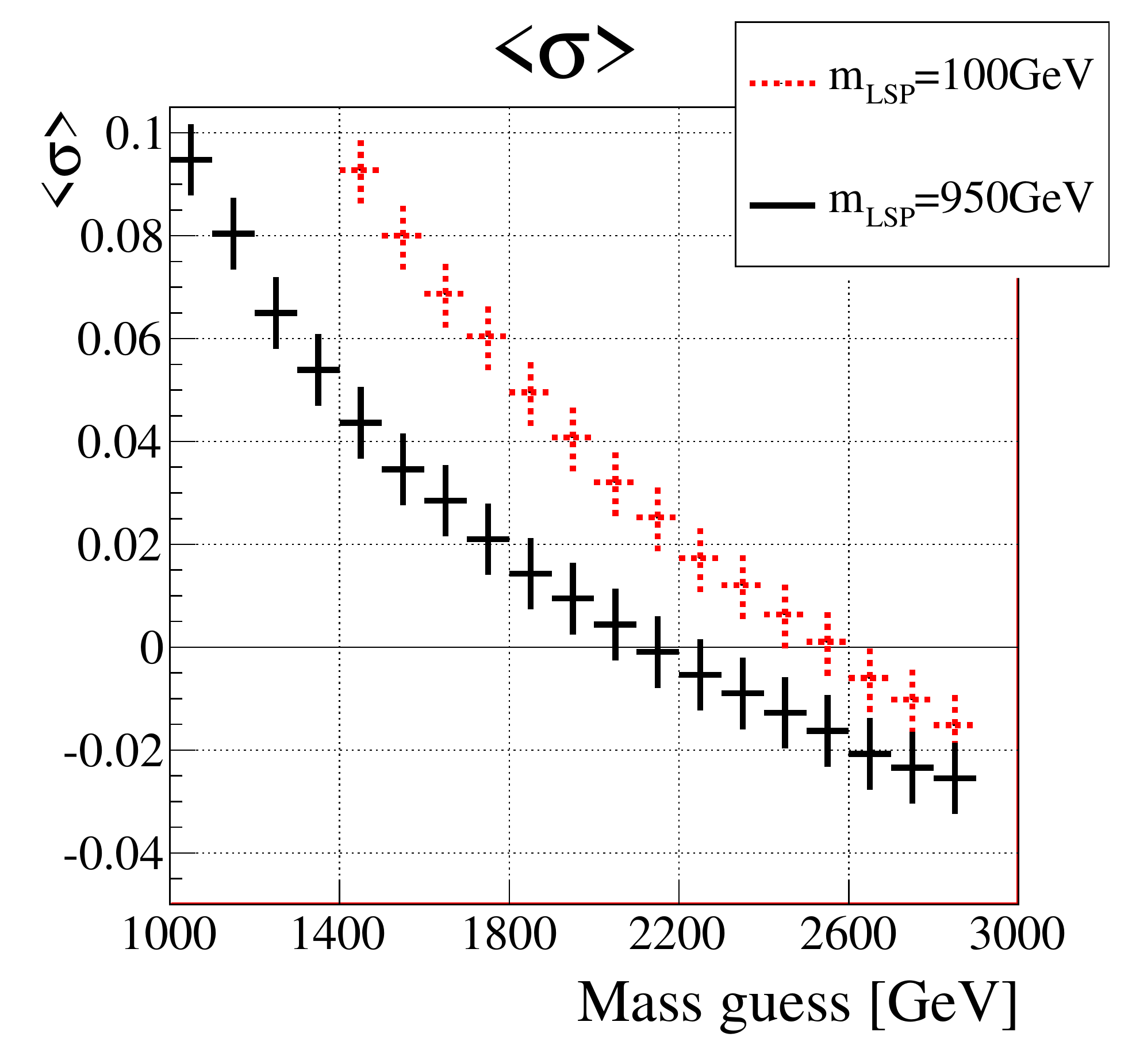}
\includegraphics[scale=0.22]{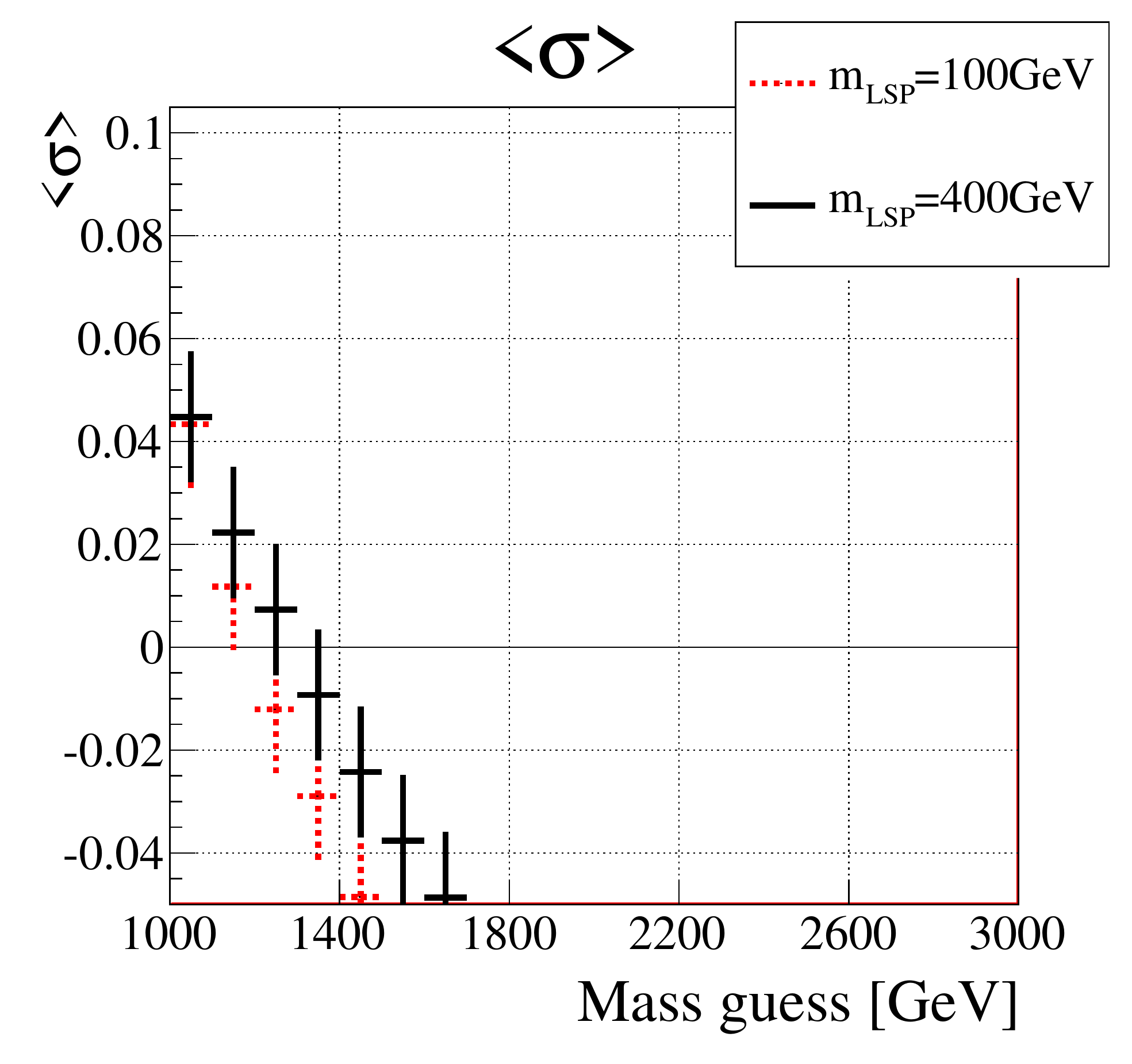}
\includegraphics[scale=0.22]{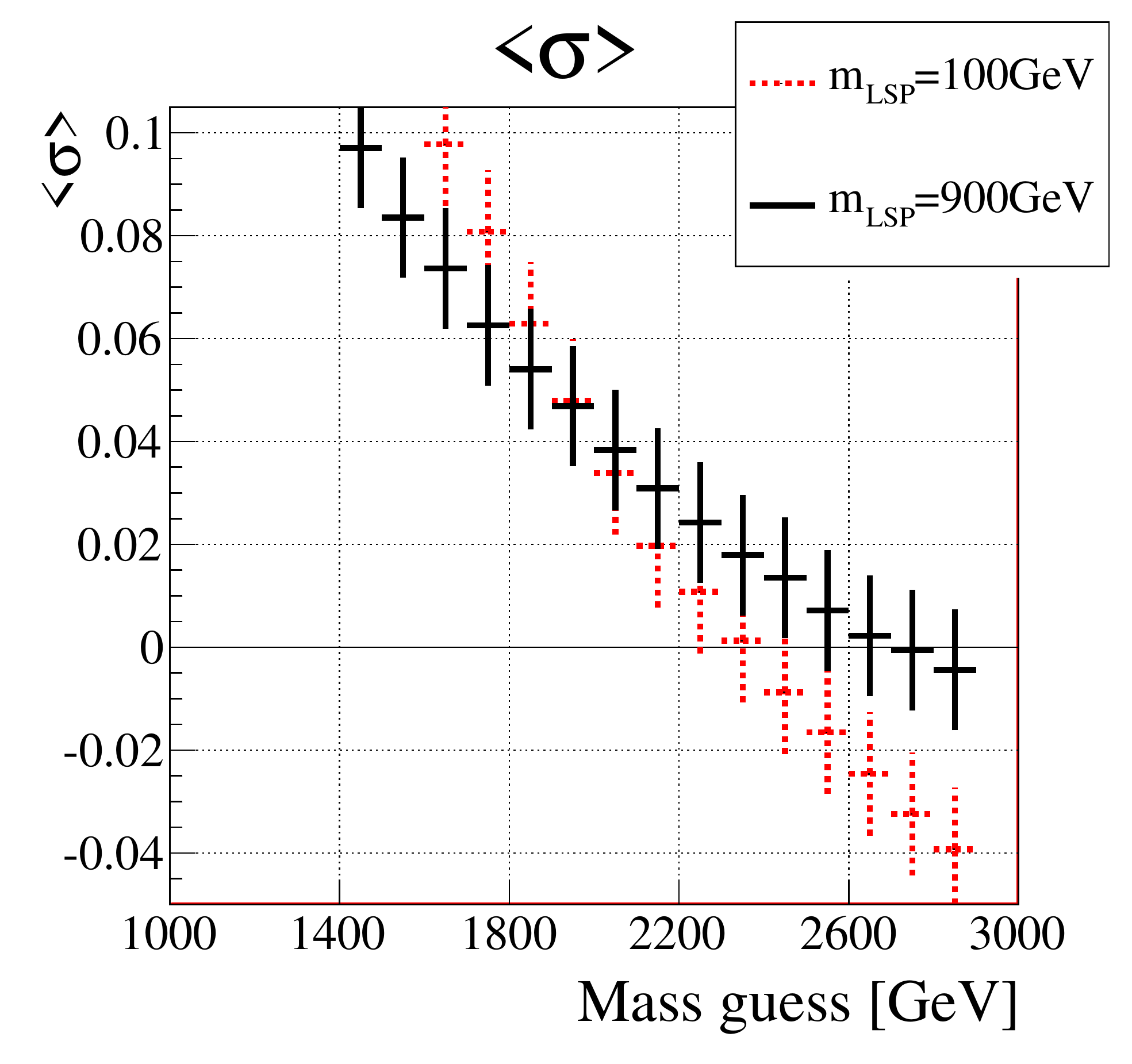}
\end{center}
\caption{The average sign of the FSR projection along the transverse ISR direction for, proceeding left to right, di-squark production
using $m_{\tilde{q}}=500~{\rm GeV}$, $m_{\tilde{q}}=1~{\rm TeV}$, and then di-gluino production with $m_{\tilde{g}}=500~{\rm GeV}$, $m_{\tilde{g}}=1~{\rm TeV}$, with the LSP mass indicated in the legends. The position at which the points intersect $\langle\sigma\rangle=0$ is what we would identify as $m_{\rm BSM}$, i.e. it where the FSR momenta are balanced
because the boost is `correct'.  We see that it is in general close to $2m_{\tilde{q}/\tilde{g}}$.  Note that the errors indicated are just the statistical errors associated with  our Monte Carlo sample sizes.  \label{fig:imb}}
\end{figure*}

\begingroup
\squeezetable
\begin{table}
\caption{ISR tagging efficiencies computed for different choices of spectra.  The first four rows
are for di-squark production, and the last four are for di-gluino production.\label{tab:isreffics} }
\begin{ruledtabular} 
\begin{tabular}{cc|cc|ccc}
\multicolumn{2}{c|}{Spectrum}   & \multicolumn{2}{c|}{Efficiencies [\%] }   & \multicolumn{3}{c}{Type  of tag applied [\%]}\\
$m_{\tilde{q}}/m_{\tilde{g}}$& $m_{\rm LSP}$& Trigger  & Mistag& \Eq{eq:pttag}& \Eq{eq:raptag} & \Eq{eq:delta}\\
\hline
500 GeV & 100 GeV & 42 & 15 & 69 & 22 & 9\\ 
500 GeV & 450 GeV & 42 & 12 & 52 & 39 & 9 \\
1 TeV & 100 GeV & 41 & 11 & 79 & 14 & 7\\ 
1 TeV & 950 GeV & 41 & 9 & 52 & 39 & 9\\ 
\hline
500 GeV & 100 GeV & 13 & 22 & 48 & 42 & 10\\ 
500 GeV & 400 GeV & 15 & 10 & 35 & 58 & 6 \\
1 TeV & 100 GeV & 12 & 25 & 59 & 30 & 11\\ 
1 TeV & 900 GeV & 16 & 8 & 37 & 57 & 6\\ 
\end{tabular}
\end{ruledtabular}
\end{table}
\endgroup
\Fig{fig:imb} shows the distribution of $\langle\sigma\rangle$ for the spectra in \Tab{tab:isreffics}  where we see that the kinematical 
technique introduced earlier works quite well: $\langle m_{\rm BSM}\rangle=1.3~\rm TeV$ for $m_{\tilde{q}}/m_{\tilde{g}}=500~{\rm GeV}$ and 
$\langle m_{\rm BSM}\rangle=2.5~\rm TeV$ for $m_{\tilde{q}}/m_{\tilde{g}}=1~{\rm TeV}$, values which are quite close to those measured by the point 
at which the FSR momentum have been boosted to have no preferred direction ($\langle\sigma\rangle=0)$.

\section{Conclusion}
\label{sec:conclusions}
While ISR is normally regarded as a nuisance, here we have seen that it can instead 
be useful, allowing for qualitatively new measurements of BSM physics that would be difficult or impossible to otherwise access.   In this paper, we have
introduced a set of techniques for tagging the ISR created in the pair production of BSM states, each of which decays into  jets and an invisible 
particle.  Although the methods we introduced are specific to this sort of process, they can be readily extended to scenarios where BSM physics 
realizes a more complicated final state topology.

We have applied these techniques to SUSY di-squark (di-gluino) production, where we saw they tagged ISR in roughly $40\%\ (15\%)$ of the events, with a mistag rate of around only $10\%\ (15\%)$.  This 
allowed us to make a qualitatively new measurement, using a kinematic technique we introduced, of the squark/gluino mass to within roughly $20\%$ of its true value, without 
making any assumptions on the rest of the SUSY spectrum. 

Given the success of this relatively simple procedure, we expect that more elaborate ISR taggers (employing, for example, the information contained in jet substructure) could be constructed for use  in other topologies to 
produce new and complementary measurements of both SM physics and physics beyond the Standard Model. 

\acknowledgments{We would like to thank D. Feldman, E. Kuflic, I. Kim, M. Lisanti,  G. Salam, M. Schwartz, I. Stewart, and J. Thaler for helpful discussions. DK is supported by a Simons postdoctoral fellowship and by an LHC-TI travel grant.  LR is supported by NSF grant PHY-0556111.  L.-T.W. is supported by the NSF under grant PHY-0756966, and by a DOE OJI award under grant DE-FG02-90ER40542}

\bibliography{isrt}

\begin{thebibliography}{37}%
\makeatletter
\providecommand \@ifxundefined [1]{%
 \@ifx{#1\undefined}
}%
\providecommand \@ifnum [1]{%
 \ifnum #1\expandafter \@firstoftwo
 \else \expandafter \@secondoftwo
 \fi
}%
\providecommand \@ifx [1]{%
 \ifx #1\expandafter \@firstoftwo
 \else \expandafter \@secondoftwo
 \fi
}%
\providecommand \natexlab [1]{#1}%
\providecommand \enquote  [1]{``#1''}%
\providecommand \bibnamefont  [1]{#1}%
\providecommand \bibfnamefont [1]{#1}%
\providecommand \citenamefont [1]{#1}%
\providecommand \href@noop [0]{\@secondoftwo}%
\providecommand \href [0]{\begingroup \@sanitize@url \@href}%
\providecommand \@href[1]{\@@startlink{#1}\@@href}%
\providecommand \@@href[1]{\endgroup#1\@@endlink}%
\providecommand \@sanitize@url [0]{\catcode `\\12\catcode `\$12\catcode
  `\&12\catcode `\#12\catcode `\^12\catcode `\_12\catcode `\%12\relax}%
\providecommand \@@startlink[1]{}%
\providecommand \@@endlink[0]{}%
\providecommand \url  [0]{\begingroup\@sanitize@url \@url }%
\providecommand \@url [1]{\endgroup\@href {#1}{\urlprefix }}%
\providecommand \urlprefix  [0]{URL }%
\providecommand \Eprint [0]{\href }%
\@ifxundefined \urlstyle {%
  \providecommand \doi  [0]{\begingroup \@sanitize@url \@doi}%
  \providecommand \@doi [1]{\endgroup \@@startlink {\doibase
  #1}doi:\discretionary {}{}{}#1\@@endlink }%
}{%
  \providecommand \doi  [0]{doi:\discretionary{}{}{}\begingroup
  \urlstyle{rm}\Url }%
}%
\providecommand \doibase [0]{http://dx.doi.org/}%
\providecommand \Doi [0]{\begingroup \@sanitize@url \@Doi }%
\providecommand \@Doi  [1]{\endgroup\@@startlink{\doibase#1}\@@Doi}%
\providecommand \@@Doi [1]{#1\@@endlink}%
\providecommand \selectlanguage [0]{\@gobble}%
\providecommand \bibinfo  [0]{\@secondoftwo}%
\providecommand \bibfield  [0]{\@secondoftwo}%
\providecommand \translation [1]{[#1]}%
\providecommand \BibitemOpen [0]{}%
\providecommand \bibitemStop [0]{}%
\providecommand \bibitemNoStop [0]{.\EOS\space}%
\providecommand \EOS [0]{\spacefactor3000\relax}%
\providecommand \BibitemShut  [1]{\csname bibitem#1\endcsname}%
\bibitem [{\citenamefont {Altarelli}\ and\ \citenamefont
  {Parisi}(1977)}]{Altarelli:1977zs}%
  \BibitemOpen
  \bibfield  {author} {\bibinfo {author} {\bibfnamefont {G.}~\bibnamefont
  {Altarelli}}\ and\ \bibinfo {author} {\bibfnamefont {G.}~\bibnamefont
  {Parisi}},\ }\Doi {10.1016/0550-3213(77)90384-4} {\bibfield  {journal}
  {\bibinfo  {journal} {Nucl. Phys.},\ }\textbf {\bibinfo {volume} {B126}},\
  \bibinfo {pages} {298} (\bibinfo {year} {1977})}\BibitemShut {NoStop}%
\bibitem [{\citenamefont {Butterworth}\ \emph {et~al.}(2008)\citenamefont
  {Butterworth}, \citenamefont {Davison}, \citenamefont {Rubin},\ and\
  \citenamefont {Salam}}]{Butterworth:2008iy}%
  \BibitemOpen
  \bibfield  {author} {\bibinfo {author} {\bibfnamefont {J.~M.}\ \bibnamefont
  {Butterworth}}, \bibinfo {author} {\bibfnamefont {A.~R.}\ \bibnamefont
  {Davison}}, \bibinfo {author} {\bibfnamefont {M.}~\bibnamefont {Rubin}}, \
  and\ \bibinfo {author} {\bibfnamefont {G.~P.}\ \bibnamefont {Salam}},\ }\Doi
  {10.1103/PhysRevLett.100.242001} {\bibfield  {journal} {\bibinfo  {journal}
  {Phys.Rev.Lett.},\ }\textbf {\bibinfo {volume} {100}},\ \bibinfo {pages}
  {242001} (\bibinfo {year} {2008})},\ \Eprint {http://arxiv.org/abs/0802.2470}
  {arXiv:0802.2470 [hep-ph]} \BibitemShut {NoStop}%
\bibitem [{\citenamefont {Ellis}\ \emph {et~al.}(2009)\citenamefont {Ellis},
  \citenamefont {Vermilion},\ and\ \citenamefont {Walsh}}]{Ellis:2009su}%
  \BibitemOpen
  \bibfield  {author} {\bibinfo {author} {\bibfnamefont {S.~D.}\ \bibnamefont
  {Ellis}}, \bibinfo {author} {\bibfnamefont {C.~K.}\ \bibnamefont
  {Vermilion}}, \ and\ \bibinfo {author} {\bibfnamefont {J.~R.}\ \bibnamefont
  {Walsh}},\ }\Doi {10.1103/PhysRevD.80.051501} {\bibfield  {journal} {\bibinfo
   {journal} {Phys.Rev.},\ }\textbf {\bibinfo {volume} {D80}},\ \bibinfo
  {pages} {051501} (\bibinfo {year} {2009})},\ \Eprint
  {http://arxiv.org/abs/0903.5081} {arXiv:0903.5081 [hep-ph]} \BibitemShut
  {NoStop}%
\bibitem [{\citenamefont {Ellis}\ \emph {et~al.}(2010)\citenamefont {Ellis},
  \citenamefont {Vermilion},\ and\ \citenamefont {Walsh}}]{Ellis:2009me}%
  \BibitemOpen
  \bibfield  {author} {\bibinfo {author} {\bibfnamefont {S.~D.}\ \bibnamefont
  {Ellis}}, \bibinfo {author} {\bibfnamefont {C.~K.}\ \bibnamefont
  {Vermilion}}, \ and\ \bibinfo {author} {\bibfnamefont {J.~R.}\ \bibnamefont
  {Walsh}},\ }\Doi {10.1103/PhysRevD.81.094023} {\bibfield  {journal} {\bibinfo
   {journal} {Phys.Rev.},\ }\textbf {\bibinfo {volume} {D81}},\ \bibinfo
  {pages} {094023} (\bibinfo {year} {2010})},\ \Eprint
  {http://arxiv.org/abs/0912.0033} {arXiv:0912.0033 [hep-ph]} \BibitemShut
  {NoStop}%
\bibitem [{\citenamefont {Krohn}\ \emph {et~al.}(2010)\citenamefont {Krohn},
  \citenamefont {Thaler},\ and\ \citenamefont {Wang}}]{Krohn:2009th}%
  \BibitemOpen
  \bibfield  {author} {\bibinfo {author} {\bibfnamefont {D.}~\bibnamefont
  {Krohn}}, \bibinfo {author} {\bibfnamefont {J.}~\bibnamefont {Thaler}}, \
  and\ \bibinfo {author} {\bibfnamefont {L.-T.}\ \bibnamefont {Wang}},\ }\Doi
  {10.1007/JHEP02(2010)084} {\bibfield  {journal} {\bibinfo  {journal} {JHEP},\
  }\textbf {\bibinfo {volume} {1002}},\ \bibinfo {pages} {084} (\bibinfo {year}
  {2010})},\ \Eprint {http://arxiv.org/abs/0912.1342} {arXiv:0912.1342
  [hep-ph]} \BibitemShut {NoStop}%
\bibitem [{\citenamefont {Alwall}\ \emph
  {et~al.}(2009){\natexlab{a}}\citenamefont {Alwall}, \citenamefont
  {Hiramatsu}, \citenamefont {Nojiri},\ and\ \citenamefont
  {Shimizu}}]{Alwall:2009zu}%
  \BibitemOpen
  \bibfield  {author} {\bibinfo {author} {\bibfnamefont {J.}~\bibnamefont
  {Alwall}}, \bibinfo {author} {\bibfnamefont {K.}~\bibnamefont {Hiramatsu}},
  \bibinfo {author} {\bibfnamefont {M.~M.}\ \bibnamefont {Nojiri}}, \ and\
  \bibinfo {author} {\bibfnamefont {Y.}~\bibnamefont {Shimizu}},\ }\Doi
  {10.1103/PhysRevLett.103.151802} {\bibfield  {journal} {\bibinfo  {journal}
  {Phys.Rev.Lett.},\ }\textbf {\bibinfo {volume} {103}},\ \bibinfo {pages}
  {151802} (\bibinfo {year} {2009}{\natexlab{a}})},\ \Eprint
  {http://arxiv.org/abs/0905.1201} {arXiv:0905.1201 [hep-ph]} \BibitemShut
  {NoStop}%
\bibitem [{\citenamefont {Nojiri}\ and\ \citenamefont
  {Sakurai}(2010)}]{Nojiri:2010mk}%
  \BibitemOpen
  \bibfield  {author} {\bibinfo {author} {\bibfnamefont {M.~M.}\ \bibnamefont
  {Nojiri}}\ and\ \bibinfo {author} {\bibfnamefont {K.}~\bibnamefont
  {Sakurai}},\ }\href@noop {} { (\bibinfo {year} {2010})},\ \Eprint
  {http://arxiv.org/abs/1008.1813} {arXiv:1008.1813 [hep-ph]} \BibitemShut
  {NoStop}%
\bibitem [{\citenamefont {Konar}\ \emph
  {et~al.}(2010){\natexlab{a}}\citenamefont {Konar}, \citenamefont {Kong},
  \citenamefont {Matchev},\ and\ \citenamefont {Park}}]{Konar:2010ma}%
  \BibitemOpen
  \bibfield  {author} {\bibinfo {author} {\bibfnamefont {P.}~\bibnamefont
  {Konar}}, \bibinfo {author} {\bibfnamefont {K.}~\bibnamefont {Kong}},
  \bibinfo {author} {\bibfnamefont {K.~T.}\ \bibnamefont {Matchev}}, \ and\
  \bibinfo {author} {\bibfnamefont {M.}~\bibnamefont {Park}},\ }\href@noop {} {
  (\bibinfo {year} {2010}{\natexlab{a}})},\ \Eprint
  {http://arxiv.org/abs/1006.0653} {arXiv:1006.0653 [hep-ph]} \BibitemShut
  {NoStop}%
\bibitem [{\citenamefont {Alwall}\ \emph
  {et~al.}(2008){\natexlab{a}}\citenamefont {Alwall}, \citenamefont {Hoeche},
  \citenamefont {Krauss}, \citenamefont {Lavesson}, \citenamefont {Lonnblad}
  \emph {et~al.}}]{Alwall:2007fs}%
  \BibitemOpen
  \bibfield  {author} {\bibinfo {author} {\bibfnamefont {J.}~\bibnamefont
  {Alwall}}, \bibinfo {author} {\bibfnamefont {S.}~\bibnamefont {Hoeche}},
  \bibinfo {author} {\bibfnamefont {F.}~\bibnamefont {Krauss}}, \bibinfo
  {author} {\bibfnamefont {N.}~\bibnamefont {Lavesson}}, \bibinfo {author}
  {\bibfnamefont {L.}~\bibnamefont {Lonnblad}},  \emph {et~al.},\ }\Doi
  {10.1140/epjc/s10052-007-0490-5} {\bibfield  {journal} {\bibinfo  {journal}
  {Eur.Phys.J.},\ }\textbf {\bibinfo {volume} {C53}},\ \bibinfo {pages} {473}
  (\bibinfo {year} {2008}{\natexlab{a}})},\ \Eprint
  {http://arxiv.org/abs/0706.2569} {arXiv:0706.2569 [hep-ph]} \BibitemShut
  {NoStop}%
\bibitem [{\citenamefont {Alwall}\ \emph {et~al.}(2010)\citenamefont {Alwall},
  \citenamefont {Freitas},\ and\ \citenamefont {Mattelaer}}]{Alwall:2010cq}%
  \BibitemOpen
  \bibfield  {author} {\bibinfo {author} {\bibfnamefont {J.}~\bibnamefont
  {Alwall}}, \bibinfo {author} {\bibfnamefont {A.}~\bibnamefont {Freitas}}, \
  and\ \bibinfo {author} {\bibfnamefont {O.}~\bibnamefont {Mattelaer}},\
  }\href@noop {} { (\bibinfo {year} {2010})},\ \Eprint
  {http://arxiv.org/abs/1010.2263} {arXiv:1010.2263 [hep-ph]} \BibitemShut
  {NoStop}%
\bibitem [{\citenamefont {Plehn}\ \emph {et~al.}(2007)\citenamefont {Plehn},
  \citenamefont {Rainwater},\ and\ \citenamefont {Skands}}]{Plehn:2005cq}%
  \BibitemOpen
  \bibfield  {author} {\bibinfo {author} {\bibfnamefont {T.}~\bibnamefont
  {Plehn}}, \bibinfo {author} {\bibfnamefont {D.}~\bibnamefont {Rainwater}}, \
  and\ \bibinfo {author} {\bibfnamefont {P.~Z.}\ \bibnamefont {Skands}},\ }\Doi
  {10.1016/j.physletb.2006.12.009} {\bibfield  {journal} {\bibinfo  {journal}
  {Phys.Lett.},\ }\textbf {\bibinfo {volume} {B645}},\ \bibinfo {pages} {217}
  (\bibinfo {year} {2007})},\ \Eprint {http://arxiv.org/abs/hep-ph/0510144}
  {arXiv:hep-ph/0510144 [hep-ph]} \BibitemShut {NoStop}%
\bibitem [{\citenamefont {Alwall}\ \emph
  {et~al.}(2009){\natexlab{b}}\citenamefont {Alwall}, \citenamefont
  {de~Visscher},\ and\ \citenamefont {Maltoni}}]{Alwall:2008qv}%
  \BibitemOpen
  \bibfield  {author} {\bibinfo {author} {\bibfnamefont {J.}~\bibnamefont
  {Alwall}}, \bibinfo {author} {\bibfnamefont {S.}~\bibnamefont {de~Visscher}},
  \ and\ \bibinfo {author} {\bibfnamefont {F.}~\bibnamefont {Maltoni}},\ }\Doi
  {10.1088/1126-6708/2009/02/017} {\bibfield  {journal} {\bibinfo  {journal}
  {JHEP},\ }\textbf {\bibinfo {volume} {0902}},\ \bibinfo {pages} {017}
  (\bibinfo {year} {2009}{\natexlab{b}})},\ \Eprint
  {http://arxiv.org/abs/0810.5350} {arXiv:0810.5350 [hep-ph]} \BibitemShut
  {NoStop}%
\bibitem [{\citenamefont {Alwall}\ \emph
  {et~al.}(2008){\natexlab{b}}\citenamefont {Alwall}, \citenamefont {Le},
  \citenamefont {Lisanti},\ and\ \citenamefont {Wacker}}]{Alwall:2008ve}%
  \BibitemOpen
  \bibfield  {author} {\bibinfo {author} {\bibfnamefont {J.}~\bibnamefont
  {Alwall}}, \bibinfo {author} {\bibfnamefont {M.-P.}\ \bibnamefont {Le}},
  \bibinfo {author} {\bibfnamefont {M.}~\bibnamefont {Lisanti}}, \ and\
  \bibinfo {author} {\bibfnamefont {J.~G.}\ \bibnamefont {Wacker}},\ }\Doi
  {10.1016/j.physletb.2008.06.065} {\bibfield  {journal} {\bibinfo  {journal}
  {Phys.Lett.},\ }\textbf {\bibinfo {volume} {B666}},\ \bibinfo {pages} {34}
  (\bibinfo {year} {2008}{\natexlab{b}})},\ \Eprint
  {http://arxiv.org/abs/0803.0019} {arXiv:0803.0019 [hep-ph]} \BibitemShut
  {NoStop}%
\bibitem [{\citenamefont {Alwall}\ \emph
  {et~al.}(2009){\natexlab{c}}\citenamefont {Alwall}, \citenamefont {Le},
  \citenamefont {Lisanti},\ and\ \citenamefont {Wacker}}]{Alwall:2008va}%
  \BibitemOpen
  \bibfield  {author} {\bibinfo {author} {\bibfnamefont {J.}~\bibnamefont
  {Alwall}}, \bibinfo {author} {\bibfnamefont {M.-P.}\ \bibnamefont {Le}},
  \bibinfo {author} {\bibfnamefont {M.}~\bibnamefont {Lisanti}}, \ and\
  \bibinfo {author} {\bibfnamefont {J.~G.}\ \bibnamefont {Wacker}},\ }\Doi
  {10.1103/PhysRevD.79.015005} {\bibfield  {journal} {\bibinfo  {journal}
  {Phys.Rev.},\ }\textbf {\bibinfo {volume} {D79}},\ \bibinfo {pages} {015005}
  (\bibinfo {year} {2009}{\natexlab{c}})},\ \Eprint
  {http://arxiv.org/abs/0809.3264} {arXiv:0809.3264 [hep-ph]} \BibitemShut
  {NoStop}%
\bibitem [{\citenamefont {Papaefstathiou}\ and\ \citenamefont
  {Webber}(2009)}]{Papaefstathiou:2009hp}%
  \BibitemOpen
  \bibfield  {author} {\bibinfo {author} {\bibfnamefont {A.}~\bibnamefont
  {Papaefstathiou}}\ and\ \bibinfo {author} {\bibfnamefont {B.}~\bibnamefont
  {Webber}},\ }\Doi {10.1088/1126-6708/2009/06/069} {\bibfield  {journal}
  {\bibinfo  {journal} {JHEP},\ }\textbf {\bibinfo {volume} {0906}},\ \bibinfo
  {pages} {069} (\bibinfo {year} {2009})},\ \Eprint
  {http://arxiv.org/abs/0903.2013} {arXiv:0903.2013 [hep-ph]} \BibitemShut
  {NoStop}%
\bibitem [{\citenamefont {Papaefstathiou}\ and\ \citenamefont
  {Webber}(2010)}]{Papaefstathiou:2010ru}%
  \BibitemOpen
  \bibfield  {author} {\bibinfo {author} {\bibfnamefont {A.}~\bibnamefont
  {Papaefstathiou}}\ and\ \bibinfo {author} {\bibfnamefont {B.}~\bibnamefont
  {Webber}},\ }\Doi {10.1007/JHEP07(2010)018} {\bibfield  {journal} {\bibinfo
  {journal} {JHEP},\ }\textbf {\bibinfo {volume} {1007}},\ \bibinfo {pages}
  {018} (\bibinfo {year} {2010})},\ \Eprint {http://arxiv.org/abs/1004.4762}
  {arXiv:1004.4762 [hep-ph]} \BibitemShut {NoStop}%
\bibitem [{Note1()}]{Note1}%
  \BibitemOpen
  \bibinfo {note} {E.g. it is not very correlated in rapidity with FSR
  jets.}\BibitemShut {Stop}%
\bibitem [{Note2()}]{Note2}%
  \BibitemOpen
  \bibinfo {note} {It is possible, although rare, for more than one jet to pass
  Eqs.(\ref {eq:pttag}-\ref {eq:delta}). If this happens we rank the tagging
  criteria by the order in which they are listed, e.g. accepting a jet which
  passes Eq.~(\ref {eq:pttag}) over one which passes only Eq.~(\ref
  {eq:delta}), etc. If more than one jet pass the same tagging criteria, we
  take as our candidate for Eq.~(\ref {eq:pttag}) the hardest jet, and for
  Eq.~(\ref {eq:delta}) the jet with the largest $\Delta $. If more than one
  jet pass Eq.~(\ref {eq:raptag}), none are accepted.}\BibitemShut {Stop}%
\bibitem [{Note3()}]{Note3}%
  \BibitemOpen
  \bibinfo {note} {We note that $\Delta $ is sensitive to higher order
  emissions, in contrast to the criteria of Eqs.(\ref {eq:pttag}-\ref
  {eq:raptag}). Further work along this line, particularly in employing jet
  substructure (see Ref.~\cite {Abdesselam:2010pt,*Salam:2009jx} for a review),
  could potentially improve upon the methods presented herein.}\BibitemShut
  {Stop}%
\bibitem [{Note4()}]{Note4}%
  \BibitemOpen
  \bibinfo {note} {We will pursue these ideas in a followup paper.}\BibitemShut
  {Stop}%
\bibitem [{\citenamefont {Hinchliffe}\ \emph {et~al.}(1997)\citenamefont
  {Hinchliffe}, \citenamefont {Paige}, \citenamefont {Shapiro}, \citenamefont
  {Soderqvist},\ and\ \citenamefont {Yao}}]{Hinchliffe:1996iu}%
  \BibitemOpen
  \bibfield  {author} {\bibinfo {author} {\bibfnamefont {I.}~\bibnamefont
  {Hinchliffe}}, \bibinfo {author} {\bibfnamefont {F.}~\bibnamefont {Paige}},
  \bibinfo {author} {\bibfnamefont {M.}~\bibnamefont {Shapiro}}, \bibinfo
  {author} {\bibfnamefont {J.}~\bibnamefont {Soderqvist}}, \ and\ \bibinfo
  {author} {\bibfnamefont {W.}~\bibnamefont {Yao}},\ }\Doi
  {10.1103/PhysRevD.55.5520} {\bibfield  {journal} {\bibinfo  {journal}
  {Phys.Rev.},\ }\textbf {\bibinfo {volume} {D55}},\ \bibinfo {pages} {5520}
  (\bibinfo {year} {1997})},\ \Eprint {http://arxiv.org/abs/hep-ph/9610544}
  {arXiv:hep-ph/9610544 [hep-ph]} \BibitemShut {NoStop}%
\bibitem [{\citenamefont {Lester}\ and\ \citenamefont
  {Summers}(1999)}]{Lester:1999tx}%
  \BibitemOpen
  \bibfield  {author} {\bibinfo {author} {\bibfnamefont {C.}~\bibnamefont
  {Lester}}\ and\ \bibinfo {author} {\bibfnamefont {D.}~\bibnamefont
  {Summers}},\ }\Doi {10.1016/S0370-2693(99)00945-4} {\bibfield  {journal}
  {\bibinfo  {journal} {Phys.Lett.},\ }\textbf {\bibinfo {volume} {B463}},\
  \bibinfo {pages} {99} (\bibinfo {year} {1999})},\ \Eprint
  {http://arxiv.org/abs/hep-ph/9906349} {arXiv:hep-ph/9906349 [hep-ph]}
  \BibitemShut {NoStop}%
\bibitem [{\citenamefont {Barr}\ \emph {et~al.}(2003)\citenamefont {Barr},
  \citenamefont {Lester},\ and\ \citenamefont {Stephens}}]{Barr:2003rg}%
  \BibitemOpen
  \bibfield  {author} {\bibinfo {author} {\bibfnamefont {A.}~\bibnamefont
  {Barr}}, \bibinfo {author} {\bibfnamefont {C.}~\bibnamefont {Lester}}, \ and\
  \bibinfo {author} {\bibfnamefont {P.}~\bibnamefont {Stephens}},\ }\Doi
  {10.1088/0954-3899/29/10/304} {\bibfield  {journal} {\bibinfo  {journal}
  {J.Phys.G},\ }\textbf {\bibinfo {volume} {G29}},\ \bibinfo {pages} {2343}
  (\bibinfo {year} {2003})},\ \Eprint {http://arxiv.org/abs/hep-ph/0304226}
  {arXiv:hep-ph/0304226 [hep-ph]} \BibitemShut {NoStop}%
\bibitem [{\citenamefont {Lester}\ and\ \citenamefont
  {Barr}(2007)}]{Lester:2007fq}%
  \BibitemOpen
  \bibfield  {author} {\bibinfo {author} {\bibfnamefont {C.}~\bibnamefont
  {Lester}}\ and\ \bibinfo {author} {\bibfnamefont {A.}~\bibnamefont {Barr}},\
  }\Doi {10.1088/1126-6708/2007/12/102} {\bibfield  {journal} {\bibinfo
  {journal} {JHEP},\ }\textbf {\bibinfo {volume} {0712}},\ \bibinfo {pages}
  {102} (\bibinfo {year} {2007})},\ \Eprint {http://arxiv.org/abs/0708.1028}
  {arXiv:0708.1028 [hep-ph]} \BibitemShut {NoStop}%
\bibitem [{\citenamefont {Cho}\ \emph {et~al.}(2008)\citenamefont {Cho},
  \citenamefont {Choi}, \citenamefont {Kim},\ and\ \citenamefont
  {Park}}]{Cho:2007qv}%
  \BibitemOpen
  \bibfield  {author} {\bibinfo {author} {\bibfnamefont {W.~S.}\ \bibnamefont
  {Cho}}, \bibinfo {author} {\bibfnamefont {K.}~\bibnamefont {Choi}}, \bibinfo
  {author} {\bibfnamefont {Y.~G.}\ \bibnamefont {Kim}}, \ and\ \bibinfo
  {author} {\bibfnamefont {C.~B.}\ \bibnamefont {Park}},\ }\Doi
  {10.1103/PhysRevLett.100.171801} {\bibfield  {journal} {\bibinfo  {journal}
  {Phys.Rev.Lett.},\ }\textbf {\bibinfo {volume} {100}},\ \bibinfo {pages}
  {171801} (\bibinfo {year} {2008})},\ \Eprint {http://arxiv.org/abs/0709.0288}
  {arXiv:0709.0288 [hep-ph]} \BibitemShut {NoStop}%
\bibitem [{\citenamefont {Cheng}\ \emph {et~al.}(2008)\citenamefont {Cheng},
  \citenamefont {Engelhardt}, \citenamefont {Gunion}, \citenamefont {Han},\
  and\ \citenamefont {McElrath}}]{Cheng:2008mg}%
  \BibitemOpen
  \bibfield  {author} {\bibinfo {author} {\bibfnamefont {H.-C.}\ \bibnamefont
  {Cheng}}, \bibinfo {author} {\bibfnamefont {D.}~\bibnamefont {Engelhardt}},
  \bibinfo {author} {\bibfnamefont {J.~F.}\ \bibnamefont {Gunion}}, \bibinfo
  {author} {\bibfnamefont {Z.}~\bibnamefont {Han}}, \ and\ \bibinfo {author}
  {\bibfnamefont {B.}~\bibnamefont {McElrath}},\ }\Doi
  {10.1103/PhysRevLett.100.252001} {\bibfield  {journal} {\bibinfo  {journal}
  {Phys.Rev.Lett.},\ }\textbf {\bibinfo {volume} {100}},\ \bibinfo {pages}
  {252001} (\bibinfo {year} {2008})},\ \Eprint {http://arxiv.org/abs/0802.4290}
  {arXiv:0802.4290 [hep-ph]} \BibitemShut {NoStop}%
\bibitem [{\citenamefont {Konar}\ \emph
  {et~al.}(2010){\natexlab{b}}\citenamefont {Konar}, \citenamefont {Kong},
  \citenamefont {Matchev},\ and\ \citenamefont {Park}}]{Konar:2009wn}%
  \BibitemOpen
  \bibfield  {author} {\bibinfo {author} {\bibfnamefont {P.}~\bibnamefont
  {Konar}}, \bibinfo {author} {\bibfnamefont {K.}~\bibnamefont {Kong}},
  \bibinfo {author} {\bibfnamefont {K.~T.}\ \bibnamefont {Matchev}}, \ and\
  \bibinfo {author} {\bibfnamefont {M.}~\bibnamefont {Park}},\ }\Doi
  {10.1103/PhysRevLett.105.051802} {\bibfield  {journal} {\bibinfo  {journal}
  {Phys.Rev.Lett.},\ }\textbf {\bibinfo {volume} {105}},\ \bibinfo {pages}
  {051802} (\bibinfo {year} {2010}{\natexlab{b}})},\ \Eprint
  {http://arxiv.org/abs/0910.3679} {arXiv:0910.3679 [hep-ph]} \BibitemShut
  {NoStop}%
\bibitem [{\citenamefont {Cohen}\ \emph {et~al.}(2010)\citenamefont {Cohen},
  \citenamefont {Kuflik},\ and\ \citenamefont {Zurek}}]{Cohen:2010wv}%
  \BibitemOpen
  \bibfield  {author} {\bibinfo {author} {\bibfnamefont {T.}~\bibnamefont
  {Cohen}}, \bibinfo {author} {\bibfnamefont {E.}~\bibnamefont {Kuflik}}, \
  and\ \bibinfo {author} {\bibfnamefont {K.~M.}\ \bibnamefont {Zurek}},\ }\Doi
  {10.1007/JHEP11(2010)008} {\bibfield  {journal} {\bibinfo  {journal} {JHEP},\
  }\textbf {\bibinfo {volume} {1011}},\ \bibinfo {pages} {008} (\bibinfo {year}
  {2010})},\ \Eprint {http://arxiv.org/abs/1003.2204} {arXiv:1003.2204
  [hep-ph]} \BibitemShut {NoStop}%
\bibitem [{\citenamefont {Stewart}\ \emph {et~al.}(2010)\citenamefont
  {Stewart}, \citenamefont {Tackmann},\ and\ \citenamefont
  {Waalewijn}}]{Stewart:2010tn}%
  \BibitemOpen
  \bibfield  {author} {\bibinfo {author} {\bibfnamefont {I.~W.}\ \bibnamefont
  {Stewart}}, \bibinfo {author} {\bibfnamefont {F.~J.}\ \bibnamefont
  {Tackmann}}, \ and\ \bibinfo {author} {\bibfnamefont {W.~J.}\ \bibnamefont
  {Waalewijn}},\ }\Doi {10.1103/PhysRevLett.105.092002} {\bibfield  {journal}
  {\bibinfo  {journal} {Phys.Rev.Lett.},\ }\textbf {\bibinfo {volume} {105}},\
  \bibinfo {pages} {092002} (\bibinfo {year} {2010})},\ \Eprint
  {http://arxiv.org/abs/1004.2489} {arXiv:1004.2489 [hep-ph]} \BibitemShut
  {NoStop}%
\bibitem [{\citenamefont {Alwall}\ \emph {et~al.}(2007)\citenamefont {Alwall},
  \citenamefont {Demin}, \citenamefont {de~Visscher}, \citenamefont {Frederix},
  \citenamefont {Herquet} \emph {et~al.}}]{Alwall:2007st}%
  \BibitemOpen
  \bibfield  {author} {\bibinfo {author} {\bibfnamefont {J.}~\bibnamefont
  {Alwall}}, \bibinfo {author} {\bibfnamefont {P.}~\bibnamefont {Demin}},
  \bibinfo {author} {\bibfnamefont {S.}~\bibnamefont {de~Visscher}}, \bibinfo
  {author} {\bibfnamefont {R.}~\bibnamefont {Frederix}}, \bibinfo {author}
  {\bibfnamefont {M.}~\bibnamefont {Herquet}},  \emph {et~al.},\ }\Doi
  {10.1088/1126-6708/2007/09/028} {\bibfield  {journal} {\bibinfo  {journal}
  {JHEP},\ }\textbf {\bibinfo {volume} {0709}},\ \bibinfo {pages} {028}
  (\bibinfo {year} {2007})},\ \Eprint {http://arxiv.org/abs/0706.2334}
  {arXiv:0706.2334 [hep-ph]} \BibitemShut {NoStop}%
\bibitem [{\citenamefont {Sjostrand}\ \emph {et~al.}(2006)\citenamefont
  {Sjostrand}, \citenamefont {Mrenna},\ and\ \citenamefont
  {Skands}}]{Sjostrand:2006za}%
  \BibitemOpen
  \bibfield  {author} {\bibinfo {author} {\bibfnamefont {T.}~\bibnamefont
  {Sjostrand}}, \bibinfo {author} {\bibfnamefont {S.}~\bibnamefont {Mrenna}}, \
  and\ \bibinfo {author} {\bibfnamefont {P.~Z.}\ \bibnamefont {Skands}},\ }\Doi
  {10.1088/1126-6708/2006/05/026} {\bibfield  {journal} {\bibinfo  {journal}
  {JHEP},\ }\textbf {\bibinfo {volume} {0605}},\ \bibinfo {pages} {026}
  (\bibinfo {year} {2006})},\ \Eprint {http://arxiv.org/abs/hep-ph/0603175}
  {arXiv:hep-ph/0603175 [hep-ph]} \BibitemShut {NoStop}%
\bibitem [{\citenamefont {Hoeche}\ \emph {et~al.}(2006)\citenamefont {Hoeche},
  \citenamefont {Krauss}, \citenamefont {Lavesson}, \citenamefont {Lonnblad},
  \citenamefont {Mangano} \emph {et~al.}}]{Hoche:2006ph}%
  \BibitemOpen
  \bibfield  {author} {\bibinfo {author} {\bibfnamefont {S.}~\bibnamefont
  {Hoeche}}, \bibinfo {author} {\bibfnamefont {F.}~\bibnamefont {Krauss}},
  \bibinfo {author} {\bibfnamefont {N.}~\bibnamefont {Lavesson}}, \bibinfo
  {author} {\bibfnamefont {L.}~\bibnamefont {Lonnblad}}, \bibinfo {author}
  {\bibfnamefont {M.}~\bibnamefont {Mangano}},  \emph {et~al.},\ }\href@noop {}
  { (\bibinfo {year} {2006})},\ \Eprint {http://arxiv.org/abs/hep-ph/0602031}
  {arXiv:hep-ph/0602031 [hep-ph]} \BibitemShut {NoStop}%
\bibitem [{\citenamefont {Cacciari}\ \emph {et~al.}()\citenamefont {Cacciari},
  \citenamefont {Salam},\ and\ \citenamefont {Soyez}}]{Cacciari:Fastjet}%
  \BibitemOpen
  \bibfield  {author} {\bibinfo {author} {\bibfnamefont {M.}~\bibnamefont
  {Cacciari}}, \bibinfo {author} {\bibfnamefont {G.}~\bibnamefont {Salam}}, \
  and\ \bibinfo {author} {\bibfnamefont {G.}~\bibnamefont {Soyez}},\
  }\href@noop {} {\enquote {\bibinfo {title} {{FastJet}},}\ }\bibinfo {note}
  {Http://fastjet.fr/}\BibitemShut {NoStop}%
\bibitem [{\citenamefont {Cacciari}\ and\ \citenamefont
  {Salam}(2006)}]{Cacciari:2005hq}%
  \BibitemOpen
  \bibfield  {author} {\bibinfo {author} {\bibfnamefont {M.}~\bibnamefont
  {Cacciari}}\ and\ \bibinfo {author} {\bibfnamefont {G.~P.}\ \bibnamefont
  {Salam}},\ }\Doi {10.1016/j.physletb.2006.08.037} {\bibfield  {journal}
  {\bibinfo  {journal} {Phys. Lett.},\ }\textbf {\bibinfo {volume} {B641}},\
  \bibinfo {pages} {57} (\bibinfo {year} {2006})},\ \Eprint
  {http://arxiv.org/abs/hep-ph/0512210} {arXiv:hep-ph/0512210} \BibitemShut
  {NoStop}%
\bibitem [{\citenamefont {Cacciari}\ \emph {et~al.}(2008)\citenamefont
  {Cacciari}, \citenamefont {Salam},\ and\ \citenamefont
  {Soyez}}]{Cacciari:2008gp}%
  \BibitemOpen
  \bibfield  {author} {\bibinfo {author} {\bibfnamefont {M.}~\bibnamefont
  {Cacciari}}, \bibinfo {author} {\bibfnamefont {G.~P.}\ \bibnamefont {Salam}},
  \ and\ \bibinfo {author} {\bibfnamefont {G.}~\bibnamefont {Soyez}},\ }\Doi
  {10.1088/1126-6708/2008/04/063} {\bibfield  {journal} {\bibinfo  {journal}
  {JHEP},\ }\textbf {\bibinfo {volume} {04}},\ \bibinfo {pages} {063} (\bibinfo
  {year} {2008})},\ \Eprint {http://arxiv.org/abs/0802.1189} {arXiv:0802.1189
  [hep-ph]} \BibitemShut {NoStop}%
\bibitem [{\citenamefont {Abdesselam}\ \emph {et~al.}(2010)\citenamefont
  {Abdesselam}, \citenamefont {Kuutmann}, \citenamefont {Bitenc}, \citenamefont
  {Brooijmans}, \citenamefont {Butterworth} \emph
  {et~al.}}]{Abdesselam:2010pt}%
  \BibitemOpen
  \bibfield  {author} {\bibinfo {author} {\bibfnamefont {A.}~\bibnamefont
  {Abdesselam}}, \bibinfo {author} {\bibfnamefont {E.}~\bibnamefont
  {Kuutmann}}, \bibinfo {author} {\bibfnamefont {U.}~\bibnamefont {Bitenc}},
  \bibinfo {author} {\bibfnamefont {G.}~\bibnamefont {Brooijmans}}, \bibinfo
  {author} {\bibfnamefont {J.}~\bibnamefont {Butterworth}},  \emph {et~al.},\
  }\href@noop {} { (\bibinfo {year} {2010})},\ \Eprint
  {http://arxiv.org/abs/1012.5412} {arXiv:1012.5412 [hep-ph]} \BibitemShut
  {NoStop}%
\bibitem [{\citenamefont {Salam}(2010)}]{Salam:2009jx}%
  \BibitemOpen
  \bibfield  {author} {\bibinfo {author} {\bibfnamefont {G.~P.}\ \bibnamefont
  {Salam}},\ }\Doi {10.1140/epjc/s10052-010-1314-6} {\bibfield  {journal}
  {\bibinfo  {journal} {Eur.Phys.J.},\ }\textbf {\bibinfo {volume} {C67}},\
  \bibinfo {pages} {637} (\bibinfo {year} {2010})},\ \Eprint
  {http://arxiv.org/abs/0906.1833} {arXiv:0906.1833 [hep-ph]} \BibitemShut
  {NoStop}%
\end{thebibliography}%
\bibliographystyle{apsrev4-1}
\end{document}